\documentclass[12pt]{article}
\usepackage{amsmath, amssymb}
\def\beq{\begin{equation}}
\def\eeq{\end{equation}}
\def\bea{\begin{eqnarray}}
\def\eea{\end{eqnarray}}
\def\bef{\begin{figure}}
\def\enf{\end{figure}}

\def\C{{\bf C}}
\def\Z{{\bf Z}}
\def\R{{\bf R}}
\def\P{{\bf P}}
\def\N{{\bf N}}

\def\CC{{\cal C}}

\def\ba{\begin{array}}
\def\ea{\end{array}}
\def\bce{\begin{center}}
\def\ece{\end{center}}

\def\vol#1{{\bf #1}}

\def\nuphb#1#2#3{Nucl. Phys. {\bf B}\vol{#1#2#3} }

\def\jgp#1#2#3{J.\ Geom.\ Phys.\ {#1} (19#2) #3}

\newcommand{\drawsquare}[2]{\hbox{%
\rule{#2pt}{#1pt}\hskip-#2pt
\rule{#1pt}{#2pt}\hskip-#1pt
\rule[#1pt]{#1pt}{#2pt}}\rule[#1pt]{#2pt}{#2pt}\hskip-#2pt
\rule{#2pt}{#1pt}}

\newcommand{\fund}{ \raisebox{-.5pt} { \drawsquare{6.5}{0.4} } }
\newcommand{\antifund}{\bar{\fund}}

\begin{document}
\begin{titlepage}
\rightline{IASSNS-HEP-00/59}
\rightline{KEK-TH-705}
\rightline{KIAS-P00055}
\rightline{BROWN-HET-1234}
\rightline{hep-th/0008091}
\vskip 1cm
\centerline{ \Large\bf{Conifolds with Discrete Torsion and Noncommutativity
}}
\vskip 1cm
\centerline{Keshav Dasgupta$^{a}$, Seungjoon Hyun$^{b}$,  Kyungho Oh$^{c}$
and Radu
Tatar$^{d}$}
\vskip 1cm
\centerline{$^a$ School of Natural Sciences, Institute for Advanced Study,
Princeton, NJ 08540, USA}
\centerline{{\tt keshav@sns.ias.edu}}
\centerline{$^b$ Theory Group, KEK, Tsukuba, Ibaraki 305-0801, Japan}
\centerline{ School of Physics, Korea Institute for Advanced Study,
Seoul 130-012, Korea}
\centerline{{\tt hyun@kias.re.kr}}
\centerline{ $^c$ Dept. of Mathematics, University of
Missouri-St. Louis,
St. Louis, MO 63121, USA }
\centerline{{\tt oh@math.umsl.edu}}
\centerline{$^d$ Dept. of Physics, Brown University,
Providence, RI 02912, USA}
\centerline{\tt tatar@het.brown.edu}
\vskip 2cm
\centerline{\bf Abstract}
We study  D3 branes at orbifolded conifold singularities in the presence
of discrete torsion. The vacuum moduli space of open strings becomes
non-commutative
due to a  deformation of the superpotential and is studied
via  the representation
theory of the moduli  algebra. It is also shown that
the center of the moduli algebra correctly describes the underlying
orbifolded conifolds.
The field theory can be obtained by a marginal deformation of the
${\cal N} = 1$ gauge theory on D3 branes at conifold singularity, the
global symmetry being broken from $SU(2) \times SU(2)$ to $U(1) \times
U(1)$. By using the AdS/CFT correspondence we argue that the marginal
deformation is related to massless KK modes of NSNS and RR two form
reduced on the compact space $T^{1,1}$. We build a $T^2$ fibration of
$T^{1,1}$ and show that a D3 brane in the bulk correspond to a D5
brane on the $T^2$ fibre. We also discuss the
possible brane construction of the system.
\def\today{\ifcase\month\or
January\or February\or March\or April\or May\or June\or
July\or August\or September\or October\or November\or December\fi,
\number\year}
\vskip 1cm
\end{titlepage}
\newpage
\section{Introduction}

D3-branes   probing  a Calabi-Yau space  provide interesting short-distance 
stringy geometry through its vacuum moduli space 
 and supersymmetric gauge field theory on their world volume. 
If the Calabi-Yau space is smooth, then the low-energy gauge field theory
is ${\cal N} = 4$
in four dimensions whose supergravity dual is
 type IIB string theory on
$AdS_5 \times S^5$ according to the
AdS/CFT correspondence~\cite{mal, mal1}. If the Calabi-Yau space becomes
singular, the story is more interesting and somewhat more
natural from the point view of the AdS/CFT correspondence. The field theory
we obtain this way has  chiral gauge sectors which is  phenomenologically
more interesting and the moduli space has a rich phase
structure which could include topologically distinct Calabi-Yau spaces
connected by flops. 

One of the most often studied case is an orbifold.
The orbifold is a space obtained by taking a quotient of a smooth space $X$
by the orbits of a discrete group $G$.
In string theory, the discrete group $G$ acts not only on the
space $X$, but also on the gauge degrees of freedom which makes the moduli
space smooth by producing so called twisted sectors.
This idea was
generalized for quotient conifolds in \cite{ura,dm,gns} and other
aspects have been studied in \cite{mp,karch,dm1,ot,kw1,pru,ura1}.

In the orbifold theory of \cite{dhvw, dhvw2}, it was observed in \cite{v}
that
there is  an ambiguity in phases chosen for twisted sector. This can be
consistently implemented in the closed string theory by
weighing the path integral sector  with an extra
phase factor.
This extra phase factor, called discrete torsion, can be classified
by the second cohomology $H^2(G,U(1))$ of the orbifold group $G$.
In the open string theory, the discrete torsion was introduced  in
\cite{do1, do2}
via projective representation of $G$ on the Chan-Paton factors and this has been
further justified and
studied in \cite{do3}. Important progress has been made towards studying of
orbifolds with
discrete torsion in \cite{fiq, vw, mr, shar, kr, le1}.

In this paper, we consider
$Z_k \times Z_k$ orbifolds of the  conifold in the presence of discrete
torsion
through the framework developed in \cite{do1}.
In contrast to
orbifold
singularities, a simple conifold singularity is a singularity in the
conformal field theory (in the absence of $B$ field) and
the singularity no longer remains in the  CFT if we have discrete
torsion \cite{vw}.

When a large number N of D3 branes are brought near the conifold
singularity, the near horizon geometry is $AdS_5\times
T^{1,1}$\cite{kw} and on the world volume of D3 branes we have a chiral
${\cal N} = 1$ $SU(N)\times SU(N)$ gauge
theory with four chiral multiplets denoted by $\phi_i$ and
$\psi_i$ ($i=1,2$) and a quartic superpotential
$$Tr(\phi_1\psi_1\phi_2\psi_2- \phi_1\psi_2\phi_2\psi_1)~.$$
For a $Z_k\times Z_l$ quotient of the conifold, the gauge symmetry of the
theory becomes
$SU(N)^{kl}\times SU(N)^{' kl}$,
the gauge degrees of freedom being realized by a certain choice of
the Chan-Paton matrices\cite{ura}. The matter is chiral thus the
choice of Chan-Paton is extremely crucial in order to avoid the anomalies.

If we consider discrete torsion, the situation is slightly different
and we 
discuss in detail in section 3. Our result is similar to the
orbifold case treated recently in \cite{le1,le2,le3}.

The models with discrete torsion are interesting from the viewpoint of
noncommutative geometry and in this case the noncommutativity is not
in the field theory itself but is realized on the
moduli space of the gauge theory. In terms of branes, the space
{\it orthogonal} to the branes becomes noncommutative. The
noncommutativity in our case is related to a
q-deformed Heisenberg algebra. Because the moduli represents a
solution of both F-term and D-term equations of field theory,
in the representations of the non-commutative algebra
only products $\phi \psi$ appear because of D-term equations.

We will also discuss massless states of the 5-dimensional supergravity
by looking at the eigenvalues of the Laplacian on $T^{1,1}$ compact
space.  In the field theory
side, the superpotential is a marginal deformation which preserves a
$U(1) \times U(1)$ global symmetry. In the supergravity side, a
marginal deformation corresponds to massless KK modes and we give an
argument that these massless KK modes come from reducing the RR and NS
forms from 10 dimensions on the compact space. This will determine a
deformation of the compact space on a $T^2$ fibration of $T^{1,1}$ and
the global symmetry $U(1) \times U(1)$ will be identified with the
symmetry on a 2-torus appearing as a fibration of $T^{1,1}$.
The identification becomes very natural in the conifold case and we
can also connect our results to those without discrete torsion.

The paper is organized as follows. In section 2 we give a detailed
mathematical description of the projective representation.
In section 3 we present a   complete discussion of the field
theory on D3 branes orthogonal to
orbifolded conifolds with discrete torsion.
In section 4 we make comments on the massless KK modes which 
appear within the AdS/CFT duality and correspond to the marginal
deformation of the superpotential.
In section 5 we speculate on the realization of some aspects of
orbifolded conifolds with discrete torsion with brane
construction.
\setcounter{equation}{0}

\section{Projective Representation and Discrete Torsion}
In \cite{dhvw, dhvw2}, string theory on an orbifold $X/G$ is
obtained by projecting out $G$ noninvariant subspace from
Hilbert space of strings on $X$ where strings are
allowed to be closed up to the action of $G$. In \cite{v}, it was discovered
that due to twisting from $G$ non-trivial phase factors
appear in the one-loop partition function of
closed string orbifold theory and they
are consistent with modular invariance
if these form a discrete torsion i.e. an element of $H^2(G, U(1))$.
When D-branes are introduced in this picture the low energy effective
field theory is constructed using quiver technique with projective
representations of the orbifold group\cite{do1,do2,do3}. Below we clarify
some
technical aspects related to projective representation and discrete torsion.

A mapping $\rho: G \to GL(n, {\bf C})$ is called a 
projective
representation of $G$ if there exists
a mapping $\alpha :  G \times  G \to U(1)$ such that
\bea
\label{proj}
\rho (g)\rho (h) = \alpha (g, h) \rho (gh)~,\nonumber \\
\rho (e) = I_n~,
\eea
for all elements $g, h \in G$, where $e$ denotes the identity element of
$\Gamma$ and $I_n$ denotes the $n \times n$ identity
matrix. The mapping $\alpha$ is called the factor system of the projective
representation $\gamma$. From the associativity of $G$, we obtain
\bea
\alpha (g, e) = \alpha (e, g) =1, \quad
\alpha (g, h)\alpha(gh,k) =\alpha (g, hk) \alpha (h, k)
\quad \forall g, h, k \in G~.
\eea
The mapping $\alpha$ satisfying these properties is called a cocyle.
The mapping $\alpha$ is called a coboundary if there is
a mapping $\gamma : G \to U(1)$  such that
$\alpha (g,h) = \gamma (g) \gamma (h) \gamma^{-1} (gh)$. Two cocyles are
equivalent if their quotient is a coboundary. The set of
equivalence classes of cocycles form a group under multiplication
which will be denoted by $H^2(G, U(1))$.
A projective representation gives rise to a central extension $E$
of $G$ by $U(1)$ and a representation $\tilde{\rho} :E \to U(n, {\bf C})$
so that the following diagram commutes:
\newline
\\
\noindent
\begin{picture}(500, 80)(0,0)
\setlength{\unitlength}{0.4mm}

\put(30, 50){$1$}
\put(90,50){$U(1)$}
\put(175, 50){$E$}
\put(255, 50){$G$}
\put(330, 50){$1$}

\put(30, 0){$1$}
\put(90,0){$U(1)$}
\put(160, 0){$GL(n, \C)$}
\put(240, 0){$PGL(n, \C)$}
\put(330, 0){$1$}

\put(40, 52){\vector(1,0){40}}
\put(40, 2){\vector(1,0){40}}
\put(120, 52){\vector(1,0){50}}
\put(120, 2){\vector(1,0){30}}
\put(195, 52){\vector(1,0){50}}
\put(205, 2){\vector(1,0){30}}
\put(275, 52){\vector(1,0){50}}
\put(290, 2){\vector(1,0){35}}

\put(100,45){\vector(0,-1){35}}
\put(180,45){\vector(0,-1){35}}
\put(260,45){\vector(0,-1){35}}

\put(85, 25){\mbox{id}}
\put(170,25){$\tilde{\rho}$}
\put(250, 25){$\hat{\rho}$}

\end{picture}
\\
\\
where $PGL(n, \C)$ is the quotient of $GL(n, \C)$ by $U(1)$ and the map
$\hat {\rho }:G \to PGL(n, \C)$ is the composition of $\rho$ with the natural
projection $GL(n, \C) \to PGL(n, \C)$.
It can be shown that there is one-to-one correspondence between
the central extensions of $G$ by $U(1)$ and the projective
representations of $G$. Two projective representations $\rho_1, \rho_2$ are
said to be
projectively equivalent if $\hat{\rho_1} = \hat{\rho_2}$.
Thus the projective representations are up to equivalences
classified  by $\mbox{Ext}(G, U(1))$ which is
equal to $H^2(G, U(1))$. Given a cocycle $\alpha$,
one can define a twisted group algebra
\bea
\C_{\alpha}G = \{\sum c_i \bar{g_i} | c_i \in \C, \,\, g_i \in G\}
\eea
with multiplication
\bea
\bar{g}\bar{h} = \alpha (g, h) \Bar{gh} \quad \forall g, h \in G.
\eea
Then there is one-to-one correspondence between $\alpha$-representations of
$G$
and $\C_{\alpha}G$-modules. For a given $\alpha$-representation
$\rho$ on the vector space $V$, $V$ becomes an
$\C_{\alpha}G$-module  via a homomorphism:
\bea
h:\C_{\alpha}G \to \mbox{End}_{\C}(V).
\eea
Hence we identify $\alpha$-representations with $\C_{\alpha}G$-modules.

In this paper, we are interested in the case $G= \Z_k \times \Z_l$.
It can be shown that \cite{kar}
\bea
H^2(\Z_k \times \Z_l, U(1)) =
\mbox{Hom} (\Z_k \otimes \Z_l, U(1)) =\Z_{p}~,
\eea
where $ p = \gcd (k,l)$.

The phase $\beta (g, h)$ appearing in the closed string theory
in the $(g,h)$-twisted sector is of  the form
\bea
\beta (g, h) = \alpha (g, h) \alpha (h,g)^{-1}~,
\eea
since $G$ is abelian.
Then $\beta $ depends only on the equivalence classes of the $\alpha$
and satisfies the following cocyle condition
\bea
\beta (g, g) =1, \quad \beta (g, h) = \beta (h,g)^{-1}, \quad
\beta (g, hk) = \beta (g,h)\beta(g, k) \quad \forall g,h,k \in G.
\eea
These properties  completely fix the $\beta$ cocycles.
Indeed,  the elements  of $H^2(\Z_k \times \Z_l, U(1))$ are of the form
\bea
\beta((a,b), (a', b')) = \omega_p^{m(ab'-a'b)}, \quad  \omega_p = e^{2\pi
i/p},
\,\, m=1, \dots ,p,
\eea
where $(a, b),  (a', b') \in \Z_k \times \Z_l$.
The different projective representations are therefore determined by
the parameter  $\epsilon = \omega_p^m$ and
we have
\bea
\rho (a,b) \rho (a',b') =  \epsilon^{-ba'} \rho (a +a', b +b')~.
\eea
Let $s$ be the smallest non-zero number such that $\epsilon^s =1$.
Any irreducible $\alpha$-representation is projectively
equivalent to
\bea
\rho (a,b) =P^aQ^b,
\eea
where
\bea
\label{PQ}
P = \mbox{diag} (1, \epsilon^{-1}, \epsilon^{-2}, \ldots
\epsilon^{-(s-1)}),\quad Q =
\begin{pmatrix}0&1&0 &\dots&0&0\\
0&0&1&\dots&0&0 \\
\vdots&\vdots&\vdots&\ddots&\vdots&\vdots\\
1&0&0&\dots&0&0
\end{pmatrix}
\eea
The number of irreducible projective representations of $G$ with cocycle
$\alpha$ equals the number of $\alpha$-regular elements of $G$.
An element $g \in G$, for abelian G, is $\alpha$-regular if
\bea
\alpha (g, h) = \alpha (h, g) \quad \forall h \in G.
\eea
Thus the number $N_\alpha$ of irreducible projective representations with
cocycle class $\alpha$ is given by
\bea
N_\alpha = \frac{1}{|G|} \sum_{g, h} \frac{\alpha (g, h)}{\alpha (h, g)}
= \frac{1}{|G|} \sum_{g, h} \beta (g, h).
\eea
All the (linearly different)
irreducible $\alpha$-representations $R_{i,j}^{\mbox {irr}}$
can be obtained by multiplying by phases:
\bea
\rho_{i,j} (1,0) = \omega_k^i \rho (1,0), \quad
\rho_{i,j} (0, 1) = \omega_l^j \rho (0,1),
\eea
where $i =0, \ldots , k/s -1,$ and $j=0, \ldots ,l/s -1$.
Since $\C_\alpha G$-modules are completely reducible, a general projective
representation $R$ is a direct sum of irreducible representations
$R_{i,j}^{\mbox {irr}}$:
\bea
R = \oplus m_{i,j} R_{i,j}^{\mbox {irr}}.
\eea
The twisted group algebra $\C_\alpha G$ itself can be regarded as
a  $\C_\alpha G$ module. The $\alpha$-representation corresponding to
$\C_\alpha G$ is called the regular $\alpha$-representation of $G$.
When $G =\Z_k \times \Z_k$, the regular representation $\C_\alpha G$ is
equal to $k R^{\mbox{irr}}$.

\setcounter{equation}{0}
\section{The orbifolded conifolds and the gauge theory of the branes }

Now we consider quotient singularities of the conifold (i.e. orbifolded
conifold). The conifold is a three dimensional hypersurface singularity in
$\C^4$ defined by:
\bea
\label{fold}
{\cal C}: \quad xy -uv = 0.
\eea
The conifold can be realized as a holomorphic quotient of $\C^4$
by the $\C^*$ action given in \cite{kw}
\bea
(A_1, A_2,B_1, B_2)\mapsto (\lambda A_1, \lambda A_2,\lambda^{-1} B_1,
\lambda^{-1} B_2)\quad\mbox{ for }\lambda \in \C^*.
\eea
Thus the charge matrix is the transpose of $Q^{'}
=(1,1,-1,-1)$ and $\Delta=\sigma$ is a convex polyhedral cone
in $\N^{'}_{\R}=\R^3$
generated by $v_1, v_2, v_3, v_4 \in \N^{'}=\Z^3$  where
\bea
v_1=(1,0,0), \quad v_2=(0,1,0),\quad  v_3=(0,0,1),\quad
v_4=(1,1,-1).
\eea
The isomorphism between the conifold ${\cal C}$ and the holomorphic
quotient is given by
\bea
\label{act}
x=A_1B_1, \quad y=A_2B_2, \quad u=A_1B_2, \quad v=A_2B_1.
\eea
We take a further quotient of the conifold ${\cal C}$ by a discrete group
$\Z_k \times \Z_l$. Here $\Z_k$ acts on $A_i, B_j$ as
\bea
\label{zk}
(A_1, A_2, B_1, B_2) \mapsto
(e^{2\pi i/k} A_1, A_2, e^{-2\pi i/k}B_1, B_2),
\eea
and $\Z_l$ acts as
\bea
\label{zl}
(A_1, A_2, B_1, B_2) \mapsto
(e^{2\pi i/l} A_1, A_2, B_1, e^{-2\pi i/l}B_2).
\eea
Thus they will act on the conifold ${\cal C}$ as
\bea
\label{xy}
(x,y,u,v) \mapsto (x, y, e^{2\pi i/k} u, e^{-2\pi i/k}v)~,
\eea
and
\bea
\label{uv}
(x,y,u,v) \mapsto (e^{2\pi i/l}x, e^{-2\pi i/l}y, u, v).
\eea
This quotient is  called the orbifolded conifold or the hyper-quotient of
the
conifold
and denoted by ${\cal C}_{kl}$. 


Consider a system of
$M$ D3 branes sitting in the
transversal direction of the orbifolded 
 conifold  in ${\R}^{1,3} \times \CC_{kk}$.
The corresponding supersymmetric gauge field theory on the world
volume of the D3 branes for the case of the conifold was
constructed by Klebanov and Witten~\cite{kw} guided by  the toric
description of
the conifold as explained above. The parameters $A_i$ and $B_j$ give rise to
the chiral superfields transforming as
$(\fund, \antifund)$ and
$(\antifund, \fund)$, respectively, with respect to the gauge group
$SU(M) \times SU (M)$. There is also an additional anomaly-free $U(1)$
R-symmetry,
under which $A_i$ and $B_j$ both have charge $1/2$.

In our case, we have orbifolded conifolds with
discrete torsion
$\epsilon = e^{2\pi i/k} \in H^2(\Z_k \times \Z_k , U(1))$.
As in \cite{ura}, we begin with $SU(k^2 M) \times SU(k^2 M)$ gauge theory
with
the  Chan-Paton degrees of freedom corresponding to
the regular representation;
\bea
R (a, b) = P^aQ^b \otimes 1_{kM}.
\eea
This breaks the gauge group to $SU(kM)^k \times SU(kM)^k$
and the gauge field projects
\bea
R(a,b) G_\mu R(a,b)^{-1} =  G_\mu ~,\label{gauge}
\eea
and the chiral superfields project according to
\bea
R(a,b) A_i R(a, b)^{-1} = (a,b)\cdot  A_i~,\nonumber \\
R(a,b )B_i R(a, b)^{-1} =  (a,b)\cdot B_i.
\eea
For the irreducible representation, the solution is of the form
\bea
\label{basic-sol}
A_1=P^{-1}Q ,\,\,
A_2=  I,\,\,
B_1= Q^{-1},\,\,
B_2= P.
\eea
Hence, for the regular representation,  the most general solution is
obtained by tensoring this $k\times k$ solution with $N\times N$ matrices.
After tensoring $(\ref{basic-sol})$ with $N \times N$ matrices and
substituting
into the $SU(kN) \times SU(kN)$ theory, we obtain an
${\cal N}=1$ supersymmetric $SU(N) \times SU(N)^{'}$
gauge theory.
The matter content is in the  following representations:
\begin{center}
\begin{tabular}{ll}
{\bf Fields} & {\bf Representations} \\
$(\phi_1)$ & $(\fund,\antifund')$ \\
$(\phi_2)$ & $(\fund,\antifund')$ \\
$(\psi_1)$ & $(\antifund, \fund')$ \\
$(\psi_2)$ & $(\antifund, \fund')$
\end{tabular}
\end{center}
The superpotential is
\bea
\label{superpot}
W \sim \mbox{Tr}{\big(} (P^{-1}Q \otimes \phi_1)( Q^{-1} \otimes \psi_1 )(I
\otimes \phi_2)
(P \otimes \psi_2)
\eea
$$-(P^{-1}Q \otimes \phi_1)
(P \otimes \psi_2)( I \otimes \phi_2)(Q^{-1} \otimes \psi_1)
{\big)} \sim \mbox{Tr}(\phi_1\psi_1\phi_2\psi_2  - \epsilon^{-1}
\phi_1\psi_2\phi_2\psi_1)~.$$

Before going any further let us comment on the global symmetries which
are preserved by this superpotential. We start with a short
discussion on some marginal deformation of the ${\cal N} = 4$ theory.
Written in an ${\cal N} = 1$ notation, the interactions are summarized
in the superpotential:
\bea
W \sim \mbox{Tr}([\Phi_1, \Phi_2] \Phi_3).
\eea
where $\Phi_i, i=1,2,3$ are the chiral multiplets components of the
${\cal N} = 4$ multiplet. In this notation, only the $U(1)_R$ symmetry of
the
${\cal N}
= 1$ supersymmetry and an $SU(3)$ that rotates the $\Phi_i$ fields are
visible. In \cite{do1,do2,le1,le2,le3} the orbifold with discrete
torsion has been discussed and the global symmetry group becomes different.
The global symmetry $SU(3)$ breaks into its Cartan subalgebra
$U(1) \times U(1)$ (the charges of one of the three fields are
determined by the charges of the other two fields) and this
together with the $U(1)_R$ symmetry will
determine a $U(1)^3$ symmetry with different phases for all the three
fields.

In our case, we compare with the theory obtained on D3 branes at a
conifold singularity \cite{kw}. The superpotential preserves a
$SU(2) \times SU(2) \times U(1)_{R}$ symmetry, the first $SU(2)$
acting on the $A_i$ fields, the second $SU(2)$ acts on the $B_i$
fields and all the $A_i, B_i$ fields have an R-symmetry charge equal
to 1/2. In the case of orbifolded conifolds with discrete torsion,
the global symmetry breaks to its Cartan subalgebra $U(1) \times U(1)$,
where the first $U(1)$ acts on  $\phi_1, \phi_2$ by
$e^{i\theta}\phi_1, e^{-i\theta}\phi_2$ and the second $U(1)$ acts on
$\psi_1, \psi_2$ by
$e^{i\theta}\psi_1, e^{-i\theta}\psi_2$.
Therefore the global symmetry of the superpotential (\ref{superpot})
is $U(1)^2 \times U(1)_{R} $.

The superpotential has conformal dimension 3 and R charge 2 so it is a
marginal deformation.
By using the AdS/CFT correspondence \cite{mal,mal1}, it should correspond to
a
massless KK mode in supergravity, and we will discuss this issue in
section 4.

After we discussed the symmetries we can go further to describe the
equations derived from it.
The F-term equation for the vacuum will be
\bea
\label{f1}
\psi_1\phi_2\psi_2- \epsilon^{-1} \psi_2\phi_2\psi_1 =0~,\nonumber \\
\psi_2\phi_1\psi_1  - \epsilon^{-1} \psi_1\phi_1\psi_2 =0~,\nonumber \\
\phi_2\psi_2\phi_1  - \epsilon^{-1} \phi_1\psi_2\phi_2 =0~,\nonumber \\
\phi_1\psi_1\phi_2 - \epsilon^{-1}\phi_2 \psi_1 \phi_1 =0~.
\eea
These relations  indicate the moduli space is non-commutative
once we introduce the discrete torsion $\epsilon$.

Note that, besides the F-term equation, we also have the D-flatness
condition
which implies that
\bea
|\phi_1|^2 + |\phi_2|^2 - |\psi_1|^2 -  |\psi_2|^2 = \zeta.
\eea

In the next section, we present  the ideas of
describing  the moduli
space following \cite{le2}.

\subsection{Noncommutative moduli space}
In \cite{le2}, the authors proposed that
one needs to express the moduli space of vacua in terms of noncommutative
algebraic
geometry to capture D-brane physics correctly. A general framework has been
developed
in \cite{connes,landi}.

Locally the moduli space can be described in terms of finitely generated
associative
algebras over $\C$ with unity. Globally, we need to glue together
the locally ringed (noncommutative) spaces constructed below.
For a given moduli algebra  ${\cal A}$, let
${\cal ZA}$ be the center of ${\cal A}$. Then ${\cal ZA}$ is
a commutative algebra and we may associate a geometric object
whose points consist of prime ideals in ${\cal ZA}$:
\bea
\mbox{Spec} {\cal ZA} =\left \{
{\mathfrak p} \,\,|\,\, {\mathfrak p} \mbox{ is a prime ideal in }
{\cal ZA}\right\}.
\eea
We can endow $\mbox{Spec} {\cal ZA}$ with a natural topology where the
smallest closed set containing a prime ideal ${\mathfrak p}$ consists of
all prime ideals containing ${\mathfrak p}$. This topology is called
Zariski topology on $\mbox{Spec} {\cal ZA}$.

Now we assume that ${\cal A}$ is a finite ${\cal ZA}$-module.
Let ${\mathfrak m}$ be a maximal ideal of ${\cal ZA}$ which is the set of
all functions vanishing at
a closed point in ${\cal ZA}$. Then
${\mathfrak m}{\cal A}$ will be two sided proper ideal in ${\cal A}$
since ${\cal A}$ is a finite ${\cal ZA}$-module. Moreover we have an
injective homomorphism
\bea
\frac{{\cal ZA}}{ {\mathfrak m} } \longrightarrow
\frac{{\cal A}} {{\mathfrak m}{\cal A}}.
\eea
Therefore ${\cal A}/{\mathfrak m}{\cal A}$ is an algebra over $\C$
which is finite dimensional
as a vector space by our assumption on finiteness. Then we look for a map
into an algebra
of $M\times M$ matrices over $\C$
\bea
\pi: \frac{{\cal A}}{{\mathfrak m}{\cal A}} \longrightarrow
\mbox{Mat}(M, \C)
\eea
whose  image will give an  irreducible representation in a sense that $\C^M$
will be
the only non-trivial space which is invariant under $\pi({\cal A}/{\mathfrak
m}{\cal A})$.
Note that the ideal ${\mathfrak m}{\cal A}$ will not be a maximal ideal of
${\cal A}$. Otherwise, the algebra ${\cal A}/{\mathfrak m}{\cal A}$ will be
an algebraic
division
algebra over ${\cal ZA}/ {\mathfrak m}\cong \C$ . Thus ${\cal A}/{\mathfrak
m}{\cal A}$ will
be isomorphic to $\C$ and all the representations will be one-dimensional,
though this is not
the case in general.
We will investigate all possible such maps and their images.
We will find that there is a unique map up to $GL(M, \C)$ conjugate
action on
$\mbox{Mat}(M, \C)$ at a generic point of ${\cal ZA}$, but
there could be many different irreducible representations at special points.
The representations will be parameterized by the space $\mbox{Spec} {\cal
ZA}$ which will be
irreducible at generic points, but
as we approach to the singular points
of the orbifolded conifolds, the representation breaks into
a direct sum of irreducible representations. Hence we may form
fractional branes.
This leads  to consider the symmetric spaces
of the moduli which is a free abelian group generated by
all possible irreducible representations arising in this manner.
The symmetric space was denoted by  ${\cal SM_{A}}$ in \cite{le2}.

\subsection{The Moduli space of the orbifolded conifolds with discrete
torsion}
In our case, the equations (\ref{f1}) show that
the vacuum moduli ${\cal M}$ is  non-commutative.
Let ${\cal M}_F$ be the vacuum moduli with
only F-term constraints. Then the
corresponding  moduli algebra  ${\cal A}_F$ is generated by
$\phi_i, \psi_i$'s. But to get gauge invariant moduli, we also
have to impose the D-term equation. The D-term equation is
given by a ${\C}^*$ action on $\phi_i, \psi_i$ i.e.
\bea
\lambda \cdot (\phi_1, \phi_2, \psi_1, \psi_2) =
(\lambda \phi_1,\lambda \phi_2, \lambda^{-1}\psi_1, \lambda^{-1}\psi_2),
\quad{\mbox for }\,\, \lambda \in \C^*~.
\eea
Hence the final moduli algebra ${\cal A}$ is generated by
${\C}^{*}$ invariant fields
$\phi_1 \psi_1$, $\psi_1 \phi_1$, $ \phi_1 \psi_2$, $\psi_2 \phi_1$,
  $\phi_2 \psi_1$,  $\psi_1 \phi_2$,   $\phi_2 \psi_2$,
$\psi_2 \phi_2$ with constraints (\ref{f1}).

To reduce the number of the generators of  ${\cal A}$ by half,
we now carefully compare the $U(1)$ R-symmetries of the conifold and the
field theory
as in \cite{kw}.  First note that the transformation which multiplies each
coordinate
by $e^{i\theta}$  acts on the canonical bundle $K$ by multiplication by
$e^{2i\theta}$.
Hence it acts on the chiral superspace (which transforms as $\sqrt{K}$)
coordinates by $e^{i\theta}$. Consider $\theta = 2\pi/k$. This gives an
element of the R-symmetry
group that acts on the conifold by $\epsilon$ and on the chiral superspace
by $\epsilon$.
On the gauge theory side, this transformation corresponds to the action
$A_i \rightarrow e^{\pi i/k} A_i$ and $B_j \rightarrow e^{\pi i/k}$ since
$A_i$ and $B_j$
have R-charge $1/2$. From the toric description, the exchange of $A_i$ and
$B_j$ is
$\Z_2$ discrete symmetry of the conifold. In the field theory side,
the exchange of $A_i$ and $B_j$ will change the superpotential
$ W =\mbox{Tr}\left(\phi_1\psi_1\phi_2\psi_2  - \epsilon^{-1}
\phi_1\psi_2\phi_2\psi_1  \right)$. To compensate this change in the
superpotential,
the R-symmetry $\Upsilon$ was  introduced in \cite{kw} which acts on chiral
superspace
coordinates by $\theta \to i \theta$, acts on gluinos by $\lambda \to
i\lambda$, leaves
invariant the superfields $A$ and $B$, and  therefore acts on femionic
components  ${\cal F}$
of $A$ or $B$ by ${\cal F} \rightarrow -i{\cal F}$. Hence we
may assume that ${\cal A}$ is generated by
\bea
\label{4gen}
\phi_1 \psi_1, \,\, \phi_1 \psi_2,\,\, \phi_2 \psi_1,\,\,\phi_2 \psi_2.
\eea
after combining the exchange of $A_i$ and $B_j$ and $\Upsilon$
transformation.
But these generators are not independent and they satisfy
\bea
\label{rel1}
\phi_1 \psi_1\phi_2 \psi_2  = \epsilon^{-1}  \phi_1 \psi_2 \phi_2 \psi_1~.
\eea
We will denote the monomial
\bea
(\phi_1 \psi_1)^{a_1} (\phi_1 \psi_2)^{a_2} (\phi_2 \psi_1)^{a_3}
(\phi_2 \psi_2)^{a_4}~,
\eea
by $(a_1, a_2, a_3, a_4)$ and the multiplication
\bea
\left[ (\phi_1 \psi_1)^{b_1} (\phi_1 \psi_2)^{b_2}
(\phi_2 \psi_1)^{b_3}
(\phi_2 \psi_2)^{b_4}\right] \cdot \left[
(\phi_1 \psi_1)^{a_1} (\phi_1 \psi_2)^{a_2} (\phi_2 \psi_1)^{a_3}
(\phi_2 \psi_2)^{a_4}\right]
\eea
by  $(b_1, b_2, b_3, b_4)\cdot(a_1, a_2, a_3, a_4)$.
Let us first study the commutative part of the moduli algebra ${\cal A}$.
The monomials
\bea
(a_1, a_2, a_3, a_4)
\eea
will be in the center ${\cal ZA}$ if and only if
\bea
\label{center}
(b_1, b_2, b_3, b_4)\cdot(a_1, a_2, a_3, a_4)
=
(a_1, a_2, a_3, a_4)\cdot(b_1, b_2, b_3, b_4)
\eea
for every monomial $(b_1,b_2, b_3,b_4)$.
Using the relation (\ref{f1}), we obtain
\bea
\label{center2}
(b_1, b_2, b_3, b_4)\cdot(a_1, a_2, a_3, a_4)
=\epsilon^{(b_2-b_3)(a_4 -a_1)-(b_4-b_1)(a_2-a_3)}
(a_1, a_2, a_3, a_4)\cdot (b_1, b_2, b_3, b_4).
\eea
Thus the monomial
$(\phi_1 \psi_1)^{a_1} (\phi_1 \psi_2)^{a_2} (\phi_2 \psi_1)^{a_3}
(\phi_2 \psi_2)^{a_4}$ is in the center iff
\bea
\label{rel2}
a_1 = a_4, a_2=a_3 \pmod{k}.
\eea
Let
\bea
x = (\phi_1\psi_1)^k, y = (\phi_2 \psi_2)^{k},
u= (\phi_1 \psi_2)^k, v = (\phi_2 \psi_1)^{k}, z = \phi_1\psi_1
\phi_2 \psi_2.
\eea
>From the relations (\ref{rel1}) and (\ref{rel2}), we can see that
the center ${\cal ZA}$ can be expressed as
\bea
xy = z^k,\,\, uv = z^k,
\eea
which is exactly the orbifolded conifold. Therefore we see
that the orbifolded conifold space is described by the commutative part of
the algebra.

After identifying the commutative center of the moduli algebra we now
consider the non-commutative points of the moduli.
Let
\bea
(x - x_0, y - y_0, u - u_0, v - v_0, z - z_0){\cal ZA}
\eea
be the maximal ideal corresponding to
a point $x=x_0, y=y_0, u=u_0, v=v_0, z= z_0$ on the orbifolded conifold,
where $z_0^k=x_0y_0 =u_0v_0$. The corresponding non-commutative points will
be given by
\bea
\frac{{\cal A}}{\left((\phi_1\psi_1)^k  -x_0, (\phi_2 \psi_2)^{k} -y_0,
(\phi_1 \psi_2)^k -u_0,
(\phi_2 \psi_1)^k-v_0,\phi_1\psi_1\phi_2 \psi_2 - z_0\right){\cal A}}.
\eea
Now we look for irreducible representations $\pi$ of
this algebra, that is, a map into a matrix algebra whose image forms an
irreducible
representation.

First we consider the most generic point
$(x_0,y_0,u_0,v_0, z_0)$ on
${\mbox{Spec}{\cal A}}$, by which we mean that none of  $x_0,u_0, v_0, y_0$
are  zero.
Then the minimal polynomials of
$\pi (\phi_1\psi_1)$ and $ \pi (\phi_2 \psi_2)$ will divide
\bea
\pi^k (\phi_1\psi_1) - x_0 =0,\,\, \pi^k (\phi_2 \psi_2) - y_0 =0
\eea
respectively.
Moreover, $\pi (\phi_1\psi_1)$ and $\pi (\phi_2 \psi_2)$
commutes. Hence there must be a common eigenvector with eigenvalues $a$ and
$d$ respectively where $a^k =x_0, \, d^k = y_0$.
We denote the common eigenvector with these eigenvalues
by
$|[a, 0, 0,d]_0>$
By acting $\pi^i (\phi_1 \psi_2),\,\, i=1, \ldots , k-1$ on
$|[a, 0, 0, d]_0>$,
we obtain  a collection of vectors
\bea
|[a, 0,0,d]_0>, |[a, 0,0,d]_1 >,
\ldots ,  |[a, 0,0,d]_{k-1} > , \nonumber \\
{\rm with} \ \ \ |[a, 0,0,d]_{i} >\equiv
\pi^i (\phi_1 \psi_2) |[a, 0, 0, d]_0>,
\eea
which are simultaneously eigenvectors of
$\pi (\phi_1\psi_1)$ with eigenvalues
$a, \epsilon a,\cdots ,\epsilon^{(k-1)}\,a$
and eigenvectors of $\pi (\phi_2 \psi_2)$ with eigenvalues
$d, \epsilon^{-1} \,d,\cdots ,\epsilon^{-(k-1)}\, d$.
A set of matrices which satisfies these conditions is
\bea
\pi (\phi_1\psi_1) =  a P^{-1}~,\,\,
\pi (\phi_1 \psi_2) = b Q^{-1}~, \,\,
\pi (\phi_2 \psi_2) = d P~,
\eea
where $P, Q$ are defined in (\ref{PQ}). It also follows  from (\ref{rel1})
that $\pi (\phi_2 \psi_1) = c Q$.
>From this construction, it is clear that it gives rise to an irreducible
representation of rank $k$ and
we denote this by $R(a,b,c,d)$. It is also clear that this
is the only possible irreducible representation up to
$GL(k, \C)$ conjugate action i.e. up to change of the basis of ${\C}^k$.
We remark that
\bea
\pi (x)  = a^k I, \,\, \pi (y)= d^k I, \,\, \pi (u) = b^k I,\,\, \pi (v)
=c^k I,\,\,
\pi (z)=ad I = \omega
bc I,
\eea
where $\omega$ is the $k$-th root of the unity and these solutions
parameterize the
orbifolded conifold generically i.e. none of the coordinates are zero.

Second we suppose $x_0$ and $u_0$ are not zero, but $y_0=v_0=0$. Then $z_0$
will be also zero and
we have $\pi (\phi_1\psi_1)$ and $\pi (\phi_1 \psi_2)$ satisfy
$\pi^k (\phi_1\psi_1) - x_0 =0$ and $\pi^k (\phi_1 \psi_2) -u_0 =0$
and are invertible.
Thus we may find an eigenvector $|a>$
of $\pi (\phi_1\psi_1)$ with eigenvalue  $a$ where $a^k = x_0$.
By acting $\pi (\phi_1 \psi_2)$ on
$|a>$ repeatedly, we obtain a set of eigenvectors
\bea
|a>, |\epsilon a>, \ldots , |\epsilon^{k-1} a>, \ \ \
{\rm where} \ \ \ |\epsilon^{i} a> \equiv \pi^{i} (\phi_1 \psi_2)
\eea
of $\pi (\phi_1\psi_1)$ with eigenvalues $a, \epsilon a, \ldots ,
\epsilon^{k-1} a$.
On the other hand, $\pi (\phi_2 \psi_2) = \pi (\phi_2 \psi_1) = 0$
since $\pi (\phi_1\psi_1)\pi (\phi_2 \psi_2) =\epsilon^{-1} \pi
(\phi_1\psi_2)\pi (\phi_2 \psi_1)=
0$ and $\pi (\phi_1\psi_1)$ , $\pi (\phi_1\psi_2)$
are invertible, hence  the representation
$R(a,b,c. d)$ remains to be irreducible i.e.
\bea
\lim_{u_0, v_0\to 0}  R(a,b,c,d) = R(a,b, 0,0)
\eea
as expected from the fact that $c=0,d=0,e=0$ is a smooth point. The other
cases with
only two coordinates zero are similar.

Now it remains to consider the singular points away from the vertex of the
orbifolded conifold , for example, $y_0 = u_0 =v_0 =z_0 =0$, but $x_0 \neq
0$.
Then $\pi (\phi_2\psi_2)$ is zero because $\pi ( \phi_1\psi_1)$ is
invertible
and $\pi ( \phi_1\psi_1)\pi (\phi_2\psi_2) =0$. It is also easy to see that
$\pi (\phi_2\psi_1)= \pi (\phi_1\psi_2) =0$ by applying $\pi (\phi_2\psi_1)$
and
$\pi (\phi_1\psi_2)$ repeatedly to an eigenvector of $\pi ( \phi_1\psi_1)$.
Since $\pi ( \phi_1\psi_1)$ satisfies the equation $\pi^k ( \phi_1\psi_1) -
x_0 =0$
and other operators are zero,
$\pi ( \phi_1\psi_1)$ decomposes into one-dimensional representations
with chracters $a, \epsilon a, \ldots , \epsilon^{k-1} a$ with $a^k = x_0$,
that is,
\bea
\label{limit}
\lim_{y_0,u_0,v_0 \to 0,} R(a,b,c,d) = R(a,0,0,0) \oplus R(\epsilon
a,0,0,0)
\oplus R(\epsilon^{n-1} a ,0,0,0).
\eea

Hence we can fractionalize
the branes along the singular locus which
contributes
new twisted states along the fixed locus of the orbifolding action $\Z_k
\times \Z_k$
on the conifold.

\setcounter{equation}{0}
\section{The AdS/CFT Correspondence}
In this section we will discuss issues concerning the near-horizon
limit for D3 branes at orbifolded conifolds with discrete torsion
singularities.

In  \cite{le1,le2,le3} a comparison was made between the field theory
marginal and relevant deformations and the corresponding deformations
of $AdS_5 \times S^5$ for a maximal supersymmetric theory.
The marginal deformations are related to massless states in the
5-dimensional supergravity \cite{kim,wi1} i.e. to the vevs for
harmonics of RR and NSNS fields. In the orbifold case, the presence of
NS and RR fields determines a non-commutative moduli space for the
D-branes and the $1-\epsilon^{-1}$ deformation correspond to background
values for the RR 3-form. In the presence of the RR background field
the D3 branes pick up a dipole moment for higher brane charge and
become extended in two additional directions \cite{my1,my2}.
In the orbifold with discrete torsion case, a D3 brane in the bulk
becomes a D5 brane wrapped on a 2-torus which is a fibration of the
five sphere.

In our case, we will observe a complex massless
scalar field obtained by reducing the two forms of
type IIB on $T^{1,1}$  by using results of
\cite{gab,raj,dal1,dal2,dal3,dal4} and this corresponds to the
marginal deformation of the field theory considered in the previous section.
Deformations which do not preserve conformality were described  in
\cite{kl1,kl2,kl3,kyra}.

We begin the search for the massless Kaluza-Klein with right
properties such that it could correspond to the marginal deformation
discussed in section 3.
In the
absence of a consistent 5D theory (by compactification on $T^{1,1}$)
it is a
very difficult problem to pin point the exact harmonic. Nevertheless,
it is
suggestive that the harmonic should be related to the complex scalar
which descends from the NSNS and RR two forms whose components are all
inside $T^{1,1}$.
To study the AdS dual of such case and similar other cases we list some
CFT operators and their corresponding AdS dual\cite{gab,dal1,dal2,dal3}:
\begin{center}
\begin{tabular}{llllll}
{\bf Operators}&{\bf $\Delta^k$}&{\bf $r$}&{\bf Multiplet}&{\bf$E_0$} &
{\bf $j,l$}\\
$Tr(\phi\psi)^k$ & ${3\over 2}k$ & $k$ & vector & ${3\over 2}k$ &
${k\over 2}$\\
$Tr(W_{\alpha}(\phi\psi)^k)$ & ${3\over 2}k+{3\over 2}$ & $k+1$ & gravitino
&
${3\over 2}k+{3\over 2}$ & ${k\over 2}$\\
$Tr(W^{\alpha}W_{\alpha}(\phi\psi)^k$ & ${3\over 2}k+3$ & $k+2$ & vector &
${3\over 2}k+3$ & ${k\over 2}$\\
$Tr(J_{\alpha{\dot\alpha}}(\phi\psi)^k)$ & ${3\over 2}k+3$ & $k$ & graviton
& ${3\over 2}k+3$ & ${k\over 2}$\\
$Tr(e^V{\bar W_{\dot\alpha}}e^{-V}(\phi\psi)^k)$ & ${3\over 2}k+{3\over 2}$
&
$k-1$ & gravitino & ${3\over 2}k+{3\over 2}$ & ${k\over 2}$\\
$Tr(e^V{\bar W_{\dot\alpha}}e^{-V}W^2(\phi\psi)^k)$ & ${3\over 2}k+
{9\over 2}$ & $k+1$ & gravitino & ${3\over 2}k+{9\over 2}$ & ${k\over 2}$
\end{tabular}
\end{center}
Here $\Delta$ is the conformal dimension, $r$ is the R-charge and $E_0$ is
the
AdS energy. As an example,
from the above identification we see that the operator
$Tr(\phi\psi)^k$ corresponds to vector multiplet
containing scalars coming from the
four form and graviton reduced on $T^{1,1}$.

In our case, from the CFT side we need chiral dilaton multiplet
of the type
\bea
\Phi^k = Tr (W^{\alpha}W_{\alpha}(\phi\psi)^k)~.
\eea
This has a conformal dimension
${3\over 2}k+3$ and R charge $k+2$\cite{dal1}. The trace is a symmetrized
trace and indices are also symmetrized $SU(2)\times SU(2)$ indices.
An important observation here
is that the quartic superpotential
$ W = \epsilon^{ij}\epsilon^{kl} Tr (\phi_i\psi_k\phi_j\psi_l)$
is not a chiral primary.
Combined with another operator $Tr (W^{\alpha}W_{\alpha})$ $-$ which is
also not a chiral primary $-$ this gives a chiral superfield which is
${\bar D}^2$ of the Konishi multiplet\cite{dal1,dal2,dal3},
\bea
\label{konishi}
\bar D \bar D K = {\bar D}^2 [Tr(\phi e^V {\bar \phi} e^{-V}) +
Tr(\psi e^V {\bar \psi} e^{-V})]~,
\eea
where $W_{\alpha}= -{1\over 4} \bar D \bar D D_{\alpha} V$, $V$ is a vector
superfield and
$\bar D$ is the
operator which annihilates a chiral superfield $S$
\bea
{\bar
  D}_{\dot\alpha}S_{(\alpha_1...\alpha_{2s_1})}(x,\theta,\bar\theta)=0~,
\eea
and $x,\theta,\bar\theta$ are the coordinate of a superspace. Therefore we
have an equation of the form
\bea
\label{cone}
Tr(W^{\alpha}W_{\alpha}) \sim {\bar D} {\bar D} K - W
\eea

The operator $Tr(W^{\alpha}W_{\alpha})$ is the one that appears in the
supergravity
spectrum and it coincides with the dilaton operator $\Phi^k$ with $k=0$.
Now we should ask
what the dilaton operator corresponds to from the supergravity
point of view. It contains a complex scalar coming from the two forms of
type IIB on $T^{1,1}$. This lies in the vector multiplet. The $j,l$ values,
which are the spin quantum numbers of the two $SU(2)$,
are given by\cite{dal1,dal2,dal3}
\bea
\label{jlval}
j = l = \vert {r-2\over 2}\vert \equiv {k\over 2}=0~.
\eea
To calculate the mass of the state we define a quantity $H_0^{-}=
H_0(j,l,r-2)$ where
$H_0(j,l,r)=6(j(j+1)+l(l+1)-{r^2\over 8})$. The mass of the state is given
by
\bea
H_0^- +1\pm 2\sqrt{H_0^-+4}~.
\eea
Recall that this gives the AdS dual of a {\it combination} of the quartic
superpotential and ${\bar D} {\bar D} K$. Therefore using this indirect
method we can infer that a background of NSNS and RR two form is
switched on.
As observed in sec. 3 for the case of
the $Z_2\times Z_2$ orbifolded conifold with $\epsilon = -1$ , the
superpotential is a marginal deformation of the quartic superpotential
of $T^{1,1}$\cite{ls}. Thus we expect the mass of the complex
scalar state should be
zero. By deforming the superpotential with ${\bar D} {\bar D} K$ kept
fixed, we will be changing the background complex scalar field. This is how
a marginal deformation of the superpotential can be related to NSNS and RR
two forms.

At this point we can compare our result to the
${\cal N} = 4$ case. In ${\cal N} = 1$ language the Konishi multiplet is
\bea
\label{koni}
K = \Phi_i e^V {\bar \Phi_i}~,
\eea
where $\Phi_i$, $i=1,2,3$, are the chiral scalars. If we denote the cubic
superpotential of ${\cal N} =4$ as $W\sim Tr([\Phi_1,\Phi_2]\Phi_3)$ then
$K$ satisfies\cite{ferzaf}
\bea
\label{kn}
{\bar D}{\bar D}K = W~.
\eea
Identification of AdS dual is now simpler.

Another interesting example, though not directly related to our work,
involving marginal deformation is given by the
operator\cite{dal4}
\bea
\label{margie}
{\cal
W}=Tr(\phi_i\psi_j\phi_k\psi_l)(\sigma^r)^{ik}(\sigma^s)^{jl}\delta_{rs}~.
\eea
This breaks the global symmetry from $SU(2)\times SU(2)$ to diagonal
$SU(2)$.
The above identification of the
dilaton multiplet and the scalar multiplet will give us the AdS dual of
${\cal W}$.

\subsection{Supergravity Duals and Mirror Symmetry}
We study  orbifolded conifold with discrete torsion in view of
the AdS/CFT correspondence. In the large $N$ limit let us first
investigate how orbifolded conifold encodes
the discrete torsion.
Note that the group $G= \Z_k \times \Z_k$
does not act freely on
\bea
T^{1,1}\equiv \{(x,y,u,v) \,|\, xy -uv =0,\,
|x|^2 + |y|^2 + |u|^2 + |v|^2 = 1\}.
\eea
The first $\Z_k$
leaves a union of two linked circles
\bea
\label{first-circle}
|x| =1, y=u=v=0, \quad |y| =1, x=u=v=0,
\eea
fixed and the second $\Z_k$ leaves
\bea
\label{second-circle}
|u| =1, x=y=v=0, \quad |v| =1, x=y=u=0,
\eea
fixed.
Along these cycles, we have locally a singularity of the type
$\C^2/\Z_k$ where $\Z_k$ is the isotropy group at the point, that is,
we will have a fibration of $A_{k-1}$ singularities along this circle
which resolves with an exceptional set of $(k-1)$ two spheres $S_i$
as noticed  in \cite{kyra}. The intersection matrix of these spheres
gives a Dynkin diagram of the $A_{k-1}$ group.
Of particular importance
       are the NSNS and RR two forms $B_{NSNS}$ and
$B_{RR}$ which give rise to the scalars:
\bea
\xi_i^{NSNS} = \int_{S_i} B_{NSNS},\quad \zeta_I^{RR} = \int_{S_i} B_{RR}
\eea
These fields are present even if we did not resolve the singularity in
the string theory because they come from the
twisted sector and they can survive in the supergravity limit if they
correspond
to
massless particles. By going around the fixed circles (\ref{first-circle})
by the first $\Z_k$
  on a closed loop,
we are actually performing a twist by the elements of the second
$\Z_k$ which don't fix
the circle. For the twisted strings that live at the orbifold circles,
going around the loop picks up a phase equal to the discrete torsion
of the cycle acting on the group element to which the twisted
state corresponds which is the discrete torsion phase in the partition
function. In our case, it sets the boundary conditions for
the massless sector states. Thus geometry differs from the standard
$T^{1,1}/G$ in that the singularities have monodromy
of the exceptional spheres.
We have the monodromies of the singularities which are located on a
fixed circle and also the fractional B field, this being
characteristics of the conifold with discrete torsion.

Now we would like to consider a deformation of $T^{1,1}$ corresponding to
a marginal deformation in our superpotential. The arguments are essentially
similar to those of \cite{le2}.
It is  observed in \cite{my1} that D3-branes in the presence of $RR$
background fields picks up a dipole moment for higher brane charge
and become extended in two additional dimensions. In our case, the AdS
dual of a marginal deformation of the field theory is turning on RR
and NS fields. In the presence of the RR field, the D3 branes become
D5-branes wrapping
two sphere which is the simplest possible configuration with
the lowest energy. In other words,
the RR background is given by the
Hodge-dual
$F_{(7)}$ which couples to a $D5$-brane, and is supported on
$\R^{1,3} \times D^3$, where $D^3$ fills in $S^2$ and we can write
$F_{(7)} = \tilde{F}_{(3)} \wedge dvol_4$. The 3-disc
$D^3$ extends along the radial direction of $AdS_5$ whose boundary is
a conformal compactification of $\R^{1,3}$. So we may write
\bea
F^{RR}_{(3)} = d\rho \wedge \tilde{C}_{(2)}.
\eea
In this case the stretching happens mostly on the radial direction.
But the Supergravity equations of motion imply that there is also a
background $H_{(3)}^{NS}$ field which does not have any component on
the AdS directions (this is contrast with \cite{kl1,kl2,kl3,kyra}
where $H_{(3)}^{NS}$ had a component on the radial direction and
determined an RG flow in the field theory. Here we preserve the
conformality). The field  $H_{(3)}^{NS}$ determines a stretching in the
compact directions, therefore there is a combined effect of
deformation, both in the radial and compact directions.
The radius of the sphere is proportional with the flux of
$\tilde{F}_{(3)}$ and $H_{(3)}^{NS}$ through the 3-disc.

We now consider
a large $k$ branch where a D3 brane at the singularity becomes a set
of  $k$ D5 branes wrapped on $k$ $S^2$
cycles which in principle could be located at different location in
the radial direction because of the dielectric effect due to the
$H_{(3)}^{RR}$ field. What happens if we bring all the $k$ 2-cycle at
the same radius and make any two of them meet at a single point?
The intersection diagram of these
$k$ spheres is exactly the extended Dynkin diagram of the  $A_{k-1}$ group.
The effect of $H_{(3)}^{RR}$ disappears so that the dielectric effect
disappears and only $H_{(3)}^{NS}$ remains. As discussed in
\cite{le2,le3}, there
are massless string modes stretching between the spherical D5
branes because the distance between neighboring spheres will become
zero and these massless string modes will determine new branches in the
moduli space which are not present for the case of non-zero RR field
when the D5 branes are frozen. An explanation for this is that D5
branes on $S^2$ cycles are fractional branes which are known to be
frozen at the singularity. In the case of aligned spherical D5 branes
at the same radius, by turning on the massless modes at the
intersections of the 2-spheres one resolves the pinched torus to a smooth
$T^2$ torus by giving sizes to the intersection points. In terms of the
representations,
the D5 brane wrapping different $S^2$ corresponds to different
one dimensional representations in (\ref{limit}). Since
the sum of $k$ irreducible representation on the singular point
is a limit of an irreducible representation of a smooth point, we may move
away
from the singular point.  Modification of the intersection points
corresponds to
move
off
a sum of fractional D5 branes from the singular points to form an integral
D3 brane
in the bulk.
This new branch of the moduli signals two torus fibration of $T^{1.1}$.
The symmetry of two torus will give rise to the global   $U(1) \times U(1)$
symmetry
of the field theory of the orbifolded conifolds with discrete torsion.
Thus the global $U(1) \times U(1)$ symmetry guides our geometric
construction of
two torus fibration of $T^{1.1}$.

Before we give an explicit description  we consider a K\"ahler deformation
of
the conifold
${\cal C}: xy - uv =0$.
We can make a K\"ahler deformation by means of blowing-up
the singular point $(0,0,0,0)$. This process will replace the singular point
with a 4 manifold  $\P^1 \times \P^1$. By rewriting
the equation (\ref{fold}) as
\bea
&&y^2(\frac{x}{y} - \frac{u}{y} \frac{v}{y}) \nonumber \\
&=& v^2 (\frac{x}{v} \frac{y}{v}- \frac{u}{v})=0
\eea
we can see that  the homogeneous coordinates of
the first $\P^1$ (resp. the second $\P^1$) is given by
$[x, y]$ (resp. $[u, v]$).
Since $T^{1,1}$ can be described as an intersection of
the conifold (\ref{fold}) with a 7-sphere
$|x|^2 + |y|^2 + |u|^2 + |v|^2 =1$.
it will be a $U(1)$ bundle over $\P^1 \times \P^1$.

Thus it is natural  to consider  two torus fibration determined by
the arguments of
$\phi_1 \psi_1/\phi_2\psi_2$ and $\phi_1 \psi_2/\phi_2\psi_1$.
Here  these correspond to the phases of  the homogeneous coordinates of the
two
$\P^1$'s in the K\"ahler deformation above. The two phases now become
$S^1$'s in
each one of the  $\P^1$'s and they will give the two torus $T^2$.
Note that the global $U(1) \times U(1)$ acts on $T^2$ freely.
In terms of explicit coordinates, as discussed in \cite{dm},
the $T^{1,1}$ can be described by an U(1)
fibration in the $x^6$ direction over $\P^1 \times \P^1$ whose basis lie
in the $(4,5,8,9)$ plane. The directions $x^4, x^8$ are taken to be
the two $S^1$ cycles inside the two $\P^1$'s respectively
and the $T^2$ fibration of $T^{1,1}$ will lie on the $x^4, x^8$ plane.
This describes a torus fibration of the $T^{1,1}$.

The AdS dual of the marginal deformation is given by turning on
the RR and the NS fields. The NS field is turned on only in compact
direction more precisely on the $(x^4, x^5, x^6)$ or $(x^6, x^8, x^9)$
directions and the RR field is turned on the $(x^4, x^5, x^7)$ or
$(x^7, x^8, x^9)$ where $x^7$ is the radial direction.

The above discussion reminds the case of an
orbifolded conifold without discrete torsion when a D3 brane
orthogonal to the singularity was a D5 brane on a torus in the
$(x^4,x^8)$ directions and the brane configuration is a Brane box
\cite{hu}.
If we now deform to the degeneration i.e. we
approach a singular circle, the torus is split into $k$ spheres. In the
case without discrete torsion they will correspond to a stripe of
boxes or a stripe of diamonds in the $x^4$ or $x^8$ directions
in a Brane Box which was discussed in \cite{kyra} to correspond
to a fractional brane. In the discrete torsion case, a fractional
brane will correspond to a D5 brane wrapped on each one of the $k$
spheres. Therefore we have a correspondence between the fractional
branes for the case without discrete torsion and with discrete
torsion.
This completes our discussion referring to the deformation of $T^{1,1}$
five dimensional space into a $T^2$ fibration.

What happens now if we make a T-duality with respect to the fibre
$T^2$? If $T^2$ has $k$ nodes and also wraps around $n$ times before
closing then it will describe an $(n,k)$ doublet of charges which
transforms under T-duality. At the singularities the  K\"ahler form of
the dual torus signals a B-field whose fractional part corresponds to
the discrete torsion phase and we will also get the monodromies at the
singularities as discussed at
the beginning  of this subsection. Therefore in the T-dual picture we
have a D3 brane orthogonal to an orbifolded conifold with discrete torsion.
So we have obtained that a T-duality on the $T^2$ fibre takes us from
the deformed $T^{1,1}$ to an orbifolded conifold with discrete torsion.

\setcounter{equation}{0}

\section{Brane configuration on non-commutative torus}
In this section we
   will compare the brane configuration from the orbifolded conifold
with and without discrete torsion.

In the absence of discrete torsion the orbifolded conifold $C_{kl}$
has a brane configuration in terms of intersecting NS5 branes to form
a brane box. The total number of free parameters in this model come from
$k+l-2$ relative positions of the branes and $kl$ intersections $-$
which blow
up to form a diamond \cite{karch}. This determines the cohomology
$h^{1,1}$ as:
$$h^{1,1} = kl + k + l -2 = (k+1)(l+1) -3~.$$
The branes configuration of  the orbifolded conifold can be described on
a torus. Once discrete torsion is introduced, the torus becomes
non-commutative. This would imply that the usual brane configurations are
meaningless here. However we could also view the non-commutative
moduli spaces as spaces with $b$ units of
background two form $B_{NSNS}$.
It turns out that with discrete torsion we also switch on a RR background
of unit $c = 1 - \epsilon ^{-1}$. Therefore the brane configuration is
classified
by
$(b,c)$ and  we need sources for $b,c$, i.e NS5 branes and Dp branes,
respectively.

A related configuration has recently been worked out in \cite{hankol}.
In their configuration the $(b,c)$ values are realized by a configuration
of orientifold planes, NS5 branes and Dp branes. The orientifold plane is
cut by both the types of branes. At the point where the NS5 branes cut the
O-plane the $b$ value jumps by N units by crossing the point
where N is the number of NS5 branes.
Similarly there is a shift of $c$ when one crosses the Dp branes. These
$b,c$ values specify the two discrete $Z_2$ charges.

When the NS5 brane is at the orientifold plane (O plane)
it can split along the O-plane as two copies of ${1\over 2}$ NS5 branes
\cite{hankol}. In general the number $n$ of ${1\over 2}$ NS5 branes is
determined by $$e^{i\int_{RP^2}~B_{NSNS}} = (-)^n~,$$ where $RP^2$ is the
space orthogonal to a O-plane.

Clearly our model should have similar kind of realization. Question is how
do
we realize the orientifold plane here? For this let us look at the $Z_2$
actions of the conifold more carefully.
\bea
\label{ztwo}
Z_2 : (x,y,u,v) \mapsto (x,y,-u,-v)~,\nonumber \\
Z_2 : (x,y,u,v) \mapsto (-x,-y,u,v)~.
\eea
Intersecting the above planes with the conifold $xy = uv$ we get a set of
fixed lines
\bea
\label{fixdline}
(x=u=v=0)~ \cup ~ (y=u=v=0) \ \ \  {\rm and} \nonumber \\
(x=y=u=0)~ \cup ~  (x=y=v=0)~.
\eea
These fixed lines are intersecting orbifold 5-planes away from the conifold
point. At the conifold point they behave effectively as orbifold 3-planes.

Under an S-duality transformation the background metric of the system will
not change and D3 brane will have strongly coupled gauge theory on its
world volume. However the orbifold 5-plane will transform into a set of
orientifold 5-plane and a D5 brane on top of each other. This is consistent
with the fact that in general an orbifold plane supports gauge fields on its
world volume. Under S-duality the combination O5-D5 will support gauge
fields\cite{sennonbps}. This is precisely how we can get orientifold
planes in our model.

Observe that we could actually start from the near horizon geometry of the
system. $AdS_5\times T^{1,1}/G$ will again have fixed orbifold 5-planes,
which under S-duality become a system of O5-D5. As discussed in
section 4, discrete torsion in the $AdS$ limit is viewed as switching on
background values of NSNS and RR three forms. This strongly suggest
that we have to invoke sources for these forms in our model. Although
the connection to \cite{hankol} is suggestive here, there are still some
points which need clarification. We will return to this in a future work.

{\bf Acknowledgments}

We would like to thank G. Dall'Agata, D.-E. Diaconescu, A. Hanany,
I. Klebanov, P. Rao and
M. Strassler for helpful discussions and  R. G. Leigh and A. Uranga
for a careful reading of the manuscript and comments.
RT would like to thank N. Seiberg and
the Institute for Advanced Study for
hospitality.
The work of KD is supported by DOE grant No. DE-FG02-90ER40542
and the work of KO is supported by NSF grant PHY-9970664.

\end{document}